\documentclass[12pt,preprint]{iopart}
\usepackage{iopams}

\newcommand{\secpard}[2]{\frac{\partial^2 #1}{\partial #2^2}}

\newcommand{\frh}{\mathfrak{h}}
\newcommand{\frg}{\mathfrak{g}}
\newcommand{\frk}{\mathfrak{k}}
\newcommand{\frl}{\mathfrak{l}}
\newcommand{\frC}{\mathcal{C}}

\begin{document}

\title{Lorentz-violating gravity and the bootstrap procedure}

\author{Michael D. Seifert}
\address{Dept.\ of Physics, Astronomy, and Geophysics, Connecticut
  College \\ 270 Mohegan Ave., New London, CT 06375, USA}
\ead{mseifer1@conncoll.edu}

\date{\today}

\begin{abstract}
  In conventional gravitational physics, the so-called ``bootstrap
  procedure'' can be used to extrapolate from a linear model of a
  rank-2 tensor to a full non-linear theory of gravity (i.e., general
  relativity) via a coupling to the stress-energy of the model.  In
  this work, I extend this procedure to a ``Lorentz-violating''
  gravitational model, in which the linear tensor field and the matter
  fields ``see'' different metrics due to a coupling between the
  tensor field and a background vector field.  The resulting model can
  be thought of as a generalized Proca theory with a non-minimal
  coupling to conventional matter.  It has a similar linearized limit
  to the better-known ``bumblebee model'', but differs at higher
  orders in perturbation theory.  Its effects are unobservable in FRW
  spacetimes, but are expected to be important in anisotropic
  cosmological spacetimes.
\end{abstract}


\section{Introduction}

Lorentz symmetry and general relativity have been intimately related
ever since their inception. The idea of Lorentz symmetry between
locally defined reference frames is inherent in the Einstein
equivalence principle, and in the description of gravity as due to the
effects of a spacetime metric \cite{Will}.

However, in recent years some physicists have started to questions
whether Lorentz symmetry is in fact an exact symmetry of nature, or
whether it could be broken and how such a breaking would manifest
itself.  One of the major frameworks for these investigations is the
\emph{Standard Model Extension} (SME) \cite{Colladay1998}, which
``extends'' the Standard Model Lagrangian by relaxing the restriction
that the operator combinations appearing in the Lagrangian be Lorentz
scalars.  The coefficients of these operators are then Lorentz
tensors, and it becomes an experimental question to measure or
constrain the components of these tensor coefficients in a particular
reference frame.  A wide variety of experiments have been performed
over the past two decades in order to constrain these coefficients
\cite{Kostelecky2008}.

Given its roots in particular physics, the picture underlying the SME
is that of fields propagating on flat spacetime.  For this reason,
research into gravitational phenomenology in the context of the SME
has almost entirely focused on the description of metric perturbations
about flat spacetime \cite{Bailey2006,Kostelecky2015,Kostelecky2016},
and almost entirely on the linearized equations of motion for these
perturbations.  (However, see \cite{Bailey2016} for a case where
second-order perturbation theory can be applied in a Lorentz-violating
gravitational context.)  In some models involving a
``Lorentz-violating'' tensor field (i.e., a tensor field whose
dynamics give it a non-zero vacuum expectation value), the
perturbations of the Lorentz-violating tensor field effectively
decouple from the linearized Einstein equation, and the linearized
Einstein equation can therefore be put into a standard form involving
the linearized Riemann tensor (and its derivatives) and various
contractions of the background value of the Lorentz-violating tensor.

These investigations provide valuable constraints on the possible
behavior of Lorentz-violating gravity models.  However, they cannot
access the full range of phenomenology that one could describe as
``Lorentz-violating gravity'', for the simple reason that they are
confined to perturbations of a flat spacetime background.  Some of the
most fascinating behavior in general relativity, as well as some of
the most sensitive constraints on it, come not from the weak-gravity
limit but from situations that cannot be viewed as ``close'' to flat
spacetime: black hole physics and cosmology.  To model these
situations, we require a full non-linear model of gravity in which
Lorentz symmetry is broken.

The question then arises how to construct such a model in a
well-motivated way.  Ideally, we would like this model to in some
sense extrapolate from the linearized Lorentz-violating gravity
picture of the SME to a fully dynamical Lorentz-violating version of
general relativity.  In the case of Lorentz-invariant gravity, there
is a known technique to make this extrapolation: the so-called
``bootstrap'' procedure.  (See
\cite{Kraichnan1955,Gupta1952,Thirring1961} among others).  One starts
with a model containing a massless symmetric rank-2 tensor field
$h_{ab}$ in flat spacetime, along with some other matter sources.  One
then adds terms to the Lagrangian that couple $h_{ab}$ to the total
stress-energy tensor of the model, including its own.  These new terms
may themselves contribute to the stress-energy tensor, so we must then
insert couplings between $h_{ab}$ and these new contributions.
Iterating this procedure generates an infinite series of terms in the
action; and the infinite series of terms involving $h_{ab}$ alone can
be shown to converge to the Einstein-Hilbert action, with $R$ being
the Ricci scalar of the metric $g_{ab} = \eta_{ab} + h_{ab}$.
Moreover, if the matter sector is not too complicated, the infinite
series of terms coupling $h_{ab}$ and the matter fields will simply
have the effect of replacing the flat spacetime matter Lagrangian with
a ``minimally coupled'' version of the matter Lagrangian, substituting
$\eta_{ab} \to g_{ab}$ and $\partial_a \to \nabla_a$.  In effect, this
procedure ``bootstraps'' a linear model into a non-linear one.

It is natural to ask whether this elegant procedure can be applied if
we relax some of the underlying assumptions.  In particular, if we
start with a linear field theory that violates Lorentz symmetry in
some way, is it still possible to apply the bootstrap procedure?  Is
there a mathematical impediment to this process?  Is the
interpretation of the resulting model the same?  In this work, I show
that the bootstrap procedure can in fact be applied even if Lorentz
symmetry is violated in the linear field theory for the tensor field
$h_{ab}$.  The result is a bimetric model, in which the Ricci
curvature appearing in the Einstein-Hilbert action is associated with
an effective metric $\tilde{g}_{ab}$ constructed in a non-linear way
from the metric $g_{ab}$ that is ``seen'' by matter and a dynamical
Lorentz-violating vector field $A_a$.

The paper is structured as follows.  In Section \ref{linear.sec}, I
will discuss what it means for a linear gravity model to be
Lorentz-violating, and how such a model can be constructed in the
presence of a background vector field.  Section \ref{bootstrap.sec}
reviews the Lorentz-invariant bootstrap procedure, and then applies it
to the Lorentz-violating models constructed in Section
\ref{linear.sec}.  Finally, Section \ref{apply.sec} briefly discusses
two simple applications of the model constructed in Section
\ref{bootstrap.sec}: the SME coefficients of the resulting model, and
the application to FRW universes.

We will use units where $c = \hbar = 1$ throughout; the sign
convention will be $(-,+,+,+)$.  Symmetrizations and
antisymmetrizations of tensors over $n$ indices will be weighted by a
factor of $1/n!$, e.g., $\nabla_{(a} A_{b)} = (\nabla_a A_b + \nabla_b
A_a)/2!$.

\section{Linearized gravity without Lorentz
  symmetry \label{linear.sec}} 

\subsection{Defining ``Lorentz violation''}

Before discussing the construction of a linear gravity model that
``violates Lorentz symmetry'', it is important to state clearly
what we mean by the phrase.  As a toy model, consider two
versions of the massless Klein-Gordon equation:
\begin{equation} \label{toyKG1}
  -\secpard{\phi}{t} + \secpard{\phi}{x} + \secpard{\phi}{y} +
  \secpard{\phi}{z} = 0
\end{equation}
\begin{equation} \label{toyKG2}
  -\secpard{\phi}{t} + \secpard{\phi}{x} + \secpard{\phi}{y} +
  \frac{1}{4} \secpard{\phi}{z} = 0
\end{equation}
Which of these equations is ``Lorentz-invariant''?  Since Lorentz
symmetry includes rotations, and since the speed of waves in the
$x$-direction and $z$-direction are different in \eref{toyKG2}, one
might conclude that only \eref{toyKG1} is Lorentz-invariant.
However, it is not hard to see that \eref{toyKG1} and \eref{toyKG2}
are equivalent if we have the freedom to redefine our coordinates; one
merely needs to rescale $z \to z/2$ in \eref{toyKG2} to obtain
\eref{toyKG1}.

More generally, given a quadratic form $\alpha^{ab}$ with signature
$(-, +, +, +)$, it is always possible to find some set of coordinates
in which the equation
\begin{equation} \label{toyKGgen}
  \alpha^{ab} \partial_a \phi \partial_b \phi = 0
\end{equation}
takes the form \eref{toyKG1}.  In such a coordinate system, the
components of $\alpha^{ab}$ will simply be the familiar components of
the (inverse) Minkowski metric $\eta^{ab}$.  This allows us to define
an ``inertial coordinate system'' to be one in which the equation of
motion for $\phi$ takes the form \eref{toyKG1}. Such sets of
coordinates are not unique, of course; the set of linear coordinate
transformations that leave the wave equation in this form will simply
be a subgroup of $GL(4)$ that is isomorphic to $SO(3,1)$, and will be
the ``Lorentz transformations'' between our inertial coordinate
systems.

In a real sense, then, it is not possible to define a ``violation'' of
Lorentz symmetry in the context of a model containing only one scalar
field obeying a wave equation.  We can always use the behavior of
this field to define our clocks and metersticks, and a preferred set
of transformations of coordinates between observers, in such a way
that the speed of wave propagation is the same in all directions and
for all observers.  Where a notion of Lorentz violation can arise is
when we have multiple fields which propagate with respect to different
metrics.  For example, if our Universe contains two massless scalar
fields $\phi$ and $\psi$, with $\phi$ propagating according to
\eref{toyKGgen} and $\psi$ propagating according to
\begin{equation} \label{toyKGgen2}
  \tilde{\alpha}^{ab} \partial_a \psi \partial_b \psi = 0
\end{equation}
with $\alpha^{ab} \neq \tilde{\alpha}^{ab}$, then generically we
cannot define a set of coordinates so that the equations of motion for
both $\phi$ and $\psi$ are both of the form \eref{toyKG1}.  (The
exception being if $\alpha^{ab} = \lambda \tilde{\alpha}^{ab}$ for
some $\lambda \neq 1$; but in this case \eref{toyKGgen2} is equivalent
to \eref{toyKGgen}.)  In mathematical terms, the $SO(3,1)$ subgroups
of $GL(4)$ which leave $\alpha^{ab}$ and $\tilde{\alpha}^{ab}$
invariant are not necessarily the same.  We are free to use either one
of these fields to define what we mean by clocks, metersticks, and
transformations between ``inertial reference frames''; but once we
have done so, the other field will ``violate Lorentz symmetry''
according to this description.

More generally, if our model contains several ``sectors'', it is
sometimes possible to define Lorentz transformations in such a way
that one of the sectors is Lorentz-invariant.  From this perspective,
it is not particularly miraculous that a ``privileged'' speed exists
in our universe; we could simply define our notion of Lorentz
transformations in such a way that the speed of light was the same for
all inertial observers.  What is remarkable, rather, is that this
privileged speed appears to be the same for all polarizations of all
fundamental fields: electromagnetic fields, fermion fields, and
gravitational fields.  Even within a sector, it is not always possible
to choose coordinates for a sector such that it becomes
Lorentz-invariant.  For example, in minimal Lorentz-violating
electrodynamics, an arbitrary Lorentz-violating Lagrangian contains
nineteen free coefficients for the photon sector, of which only nine
can be shifted to the matter sector \cite{Bailey2004}.  The remaining
ten coefficients cause light to have a polarization-dependent speed
(i.e., birefringence), and so cannot be removed via a simple
coordinate redefinition.  Similar redefinitions can shift nine of the
coordinates in the gravitational SME into the photon sector or vice
versa \cite{Kostelecky2011}.

In the context of this work, we will assume that this choice has
already been made in some portion of the ``matter sector'', which
propagates according to some metric $\eta^{ab}$; our notion of Lorentz
transformations will be those transformations which leave this matter
metric unchanged. I will call this metric the \emph{fiducial metric}.
I will remain agnostic as to whether all parts of the matter sector
propagate according to the fiducial metric, though the simplest choice
(see below) would be that all of them do.

\subsection{Constructing the propagator \label{propagator.sec}}

Assuming that we have defined a fiducial metric and a notion of
Lorentz symmetry with reference to some portion of the matter sector,
the question then arises what sorts of phenomenology can arise if the
``linearized gravity sector'' is not Lorentz-symmetric.  In general,
one can imagine two broad classes of Lorentz-violating effects:
direction-dependent propagation speeds, and polarization-dependent
propagation speeds (i.e., ``gravitational birefringence''.)  While
this question was addressed in a group-theoretic context in
\cite{Kostelecky2016}, it is instructive to take an axiomatic
approach tothis matter: if we make a certain set of assumptions
about the propagation of the linearized gravity field, which types of
effects are allowed?

In choosing my axioms, I will adopt a philosophy of \emph{ceteris
  paribus}: I will attempt to construct a model that preserves as many
key features of conventional linearized gravity as possible, while
relaxing the assumption of Lorentz symmetry.  These features include:
\begin{enumerate}
\item Being described by a rank-2 symmetric tensor $h_{ab}$; \label{tensor.crit}
\item Being expressible in terms of an action principle; \label{action.crit}
\item Having second-order questions of motion; \label{secorder.crit}
\item Being coupled to a conserved stress-energy tensor. \label{cons.crit}
\end{enumerate}
Under criteria \ref{tensor.crit}, \ref{action.crit}, and
\ref{secorder.crit}, the Lagrange density for the free field $h_{ab}$
must be of the form
\begin{equation} \label{lin.lag} \mathcal{L} = \frac{1}{2} \left[
    \mathcal{P}^{abcdef} \partial_a h_{bc} \partial_d h_{ef} +
    \mathcal{R}^{abcef} (\partial_a h_{bc}) h_{ef} +
    \mathcal{Q}^{bcef} h_{bc} h_{ef} \right]
\end{equation}
for some tensors $\mathcal{P}^{abcdef}$, $\mathcal{R}^{abcde}$, and
$\mathcal{Q}^{abcd}$.  These tensors are assumed to be constant in
spacetime, but they will in general involve some additional
``background geometry'': they should not be expected to remain
invariant under the Lorentz transformations that keep $\eta^{ab}$
fixed.

From \eref{lin.lag}, we can see that $\mathcal{P}^{abcdef}$ can be
taken to be symmetric under the simple exchanges $b \leftrightarrow c$
and $e \leftrightarrow f$, and under the simultaneous exchange $\{abc
\} \leftrightarrow \{def \}$.  Similarly, $\mathcal{Q}^{bcef}$ can be
taken to be symmetric under the exchanges $b \leftrightarrow c$, $e
\leftrightarrow f$, and $\{bc \} \leftrightarrow \{ef \}$.  Finally,
since we can write
\begin{equation}
  \left(\mathcal{R}^{abcef} + \mathcal{R}^{aefbc} \right) (\partial_a
  h_{bc} ) h_{ef} = \partial_a \left( \mathcal{R}^{abcef} h_{bc}
    h_{ef} \right),
\end{equation}
it follows that the part of $\mathcal{R}^{abcef}$ that is symmetric
under the exchange $\{bc \} \leftrightarrow \{ef\}$ does not
contribute to the equations of motion.  We can thus take
$\mathcal{R}^{abcef}$ to be antisymmetric under the exchange
$\{bc \} \leftrightarrow \{ef\}$, as well as being symmetric under the
exchanges $b \leftrightarrow c$ and $e \leftrightarrow f$.

The equations of motion that arise from \eref{lin.lag} are
\begin{equation} \label{lin.EOM} - \mathcal{P}^{abcdef} \partial_a
  \partial_d h_{ef} - \mathcal{R}^{abcef} \partial_a h_{ef} +
  \mathcal{Q}^{bcef} h_{ef} = 0,
\end{equation}
or, in momentum space,
\begin{equation} \label{dispersion.rel}
  \mathcal{P}^{abcdef} k_a k_d h_{ef} - i \mathcal{R}^{abcef} k_a
  h_{ef} + \mathcal{Q}^{bcef} h_{ef} = 0,
\end{equation}
We note from this equation that we can take $\mathcal{P}^{abcdef}$ to
be symmetric under the exchange $a \leftrightarrow d$.

We can now apply criterion \ref{cons.crit} to this equation.  We will
eventually want to couple \eref{lin.EOM} to the matter stress-energy
tensor $T^{bc}$.  In the linearized limit about flat spacetime, we
expect this tensor to be identically conserved:
$\partial_b T^{bc} = 0$.  This implies that the divergence of
\eref{lin.EOM} must also vanish identically;  in momentum space, this
means that
\begin{equation} \label{contraction.id}
  \mathcal{P}^{abcdef} k_a k_b k_d -  i \mathcal{R}^{abcef} k_a k_b +
  \mathcal{Q}^{bcef} k_b  = 0
\end{equation}
for any four-vector $k_b$. Note that given the symmetries of
$\mathcal{P}^{abcdef}$, $\mathcal{R}^{abcef}$, and
$\mathcal{Q}^{bcef}$, the condition \eref{contraction.id} is
equivalent to \eref{lin.EOM} being invariant under the customary gauge
transformation $h_{ab} \to h_{ab} + \partial_{(a} \xi_{b)}$.

In principle, any set of tensors $\mathcal{P}^{abcdef}$,
$\mathcal{R}^{abcef}$, and $\mathcal{Q}^{bcef}$ with the appropriate
symmetries and satisfying \eref{contraction.id} would provide a
Lorentz-violating equation of motion for $h_{ab}$.  However, the
underlying picture we have is that this tensor is due to a coupling
between $h_{ab}$ and some new fundamental field that spontaneously
breaks Lorentz symmetry.  The simplest choice for such a field is a
Lorentz vector $A_a$; the propagator tensor $\mathcal{P}^{abcdef}$
must then be constructed locally out of tensor products of $A_a$ and
the fiducial metric $\eta^{ab}$.  The question then becomes how many
distinct tensors there are which can be so constructed and which
satisfy the desired symmetry properties and the contraction identity
\eref{contraction.id}.

To answer this question, we can simply write down a list of all
tensors of a given rank, without any particular symmetry properties,
that can be so constructed.  The most general such tensor of a given
rank must be a linear combination of these; and the symmetry
requirements and the contraction identity \eref{contraction.id} will
then place constraints on the coefficients of each tensor in this
linear combination.  For example, suppose we want to construct a
rank-6 tensor from the metric alone. such a tensor must be constructed
from three ``copies'' of the metric, and so the six indices must be
paired off; there are fifteen such pairings.  The most general rank-6
tensor that can be constructed from the metric is then
\begin{equation}
  C_1 \eta^{ab} \eta^{cd} \eta^{ef} + C_2 \eta^{ab} \eta^{ce}
  \eta^{df} + C_3 \eta^{ab} \eta^{cf} \eta^{de} + \dots,
\end{equation}
where the $C_i$ are arbitrary coefficients.  However, if we require
that this expression be symmetric under the exchanges
$b \leftrightarrow c$, $e \leftrightarrow f$, and
$\{abc\} \leftrightarrow \{def\}$, and require it to obey the
contraction identity \eref{contraction.id}, then it is
straightforward (if a bit tedious) to show that there is only one such
tensor:
\begin{eqnarray} \label{prop.LI}
  \fl \mathcal{P}_{(0)}^{abcdef} = \eta^{a(b} \eta^{c)d} \eta^{ef} +
  \eta^{a(e} \eta^{f)d}
  \eta^{bc} - \eta^{a(b} \eta^{c)(e} \eta^{f)d} \nonumber \\
  {} - \eta^{a(e} \eta^{f)(b} \eta^{c)d} - \eta^{ad} \eta^{bc}
  \eta^{ef} + \eta^{ad} \eta^{b(e} \eta^{f)c}
\end{eqnarray}
This expression, when contracted with $\partial_a \partial_d h_{ef}$
as in \eref{lin.EOM}, yields the standard linearized Einstein
equation.

Similarly, there are 45 tensors that can be constructed from two
copies of the metric and two copies of the vector $A^a \equiv
\eta^{ab} A_b$; and it is also straightforward (if rather more
tedious) to show that there is also only one possible combination of
them that satisfies the desired symmetries and the contraction
identity:
\begin{eqnarray}
  \fl \mathcal{P}_{(1)}^{abcdef} = \eta^{a(b} \eta^{c)d} A^e A^f +
  \eta^{a(e} \eta^{f)d} A^b A^c - 2 A^{(a} \eta^{d)(b}
  \eta^{c)(e} A^{f)} - 2 A^{(a} \eta^{d)(e} \eta^{f)(b} A^{c)}
  \nonumber \\ {} -
  2 A^{(b} \eta^{c)(a} \eta^{d)(e} A^{f)} + 2 \eta^{bc} A^{(a}
  \eta^{d)(e} A^{f)} + 2 \eta^{ef} A^{(a} \eta^{d)(b} A^{c)} + 2
  \eta^{ad} A^{(b} \eta^{c)(e} A^{f)} \nonumber \\ {} - \eta^{ad} \eta^{bc} A^e
  A^f - \eta^{ad} \eta^{ef} A^b A^c - \eta^{bc} \eta^{ef} A^a A^d
  + \eta^{b(e} \eta^{f)c} A^a A^d.
\end{eqnarray}
There are fifteen rank-6 tensors that can be constructed from four
copies of $A^a$ and one copy of the metric;  however, it can be shown
via similar techniques that any linear combination of these tensors
with the desired symmetry properties must vanish.  The tensor $A^a A^b
A^c A^d A^e A^f$ must also be excluded from our expression for
$\mathcal{P}^{abcdef}$ by a similar logic.

We can also apply the same logic to the rank-5 tensor
$\mathcal{R}^{abcef}$ and the rank-4 tensor $\mathcal{Q}^{bcef}$; when
we do, however, we find that these tensors must vanish.  This implies
that the dispersion relation \eref{dispersion.rel} for wave solutions
is homogeneous, i.e., if a plane wave of the form
\begin{equation}
  h_{ab} (x^c) = h^{(0)}_{ab} e^{i k_c x^c}
\end{equation}
is a solution of the equations of motion, then it remains a solution
under the substitution $k_\mu \to \lambda k_\mu$ for any $\lambda$.
This ensures (among other things) that the speed of a wave is
independent of its frequency for a given polarization and a given
direction of propagation.

Since the Lagrange density is only defined up to an overall factor,
this means that the most general possible expression for our
propagator tensor is
\begin{equation} \label{genprop}
  \mathcal{P}^{abcdef} = \mathcal{P}_{(0)}^{abcdef} + \xi
  \mathcal{P}_{(1)}^{abcdef},
\end{equation}
where $\xi$ is a free parameter.  However, it is not hard to show
(albeit, again, tedious) that this expression is equivalent to taking
the Lorentz-symmetric graviton propagator \eref{prop.LI} and
substituting
\begin{equation} \label{effmetricdef}
  \eta^{ab} \to \tilde{\eta}^{ab} \equiv \eta^{ab} + \xi A^a A^b.
\end{equation}
In other words: in the presence of a background vector field $A_a$,
and assuming the criteria listed above, the only modification to the
linearized Einstein equation that is possible is to change the
effective metric that governs the propagation of the waves and their
polarization states. (The usual ``transverse traceless gauge'' for
these waves would be defined via $\tilde{\eta}^{ab} \partial_a h_{bc}
= 0$ and $h \equiv \tilde{\eta}^{ab} h_{ab} = 0$, rather than the
equivalent expressions with $\eta^{ab}$.)  It is not possible to
define a propagator which allows for ``gravitational birefringence'',
i.e., a polarization-dependent speed of gravity.  This is consistent
with the results found in \cite{Kostelecky2016}; criterion
\ref{secorder.crit} above restricts us to what were called ``$d = 4$
operators'' in that work.

\section{Bootstrapping Lorentz-violating linearized
  gravity \label{bootstrap.sec}} 

\subsection{Deser bootstrap procedure for conventional
  gravity \label{LI.Deser.sec}} 

In Section \ref{propagator.sec}, we found that under certain
assumptions, the only way to modify the linearized Einstein equation
to include a coupling to a ``Lorentz-violating'' vector field $A_a$ is
by modifying the metric that appears in the Lorentz-invariant equation
of motion for $h_{ab}$ \eref{prop.LI}.  This modification of the
linearized Einstein equation is, in itself, self-consistent.  However,
we also know that it is possible to extend Lorentz-invariant
linearized gravity to a non-linear theory (namely, conventional
general relativity) by self-consistently coupling $h_{ab}$ to all
sources of stress-energy in the theory, including its own.  The
question then arises whether a similar procedure can be applied to a
model in which $h_{ab}$'s propagation is governed by the effective
metric \eref{effmetricdef}, or whether there is an impediment to
this.

To frame this discussion, it will help to first review the bootstrap
procedure proposed by Deser \cite{Deser1970}.  (See also
\cite{Kostelecky2009} for a more detailed explanation of this
procedure.)  In this procedure, the fundamental fields are a tensor
density $\frh^{ab}$ and a rank-(1,2) undensitized tensor
$C^{a} {}_{bc} = C^a {}_{(bc)}$.  The linear Lagrange
density is written in first-order form:
\begin{equation}
  \mathcal{L} = \mathcal{L}_{G,lin}^{LI} + \mathcal{L}_\mathrm{mat} [
  \eta, \Psi^A ],
\end{equation}
where 
\begin{equation} \label{lag.LI.lin.Pal}
  \mathcal{L}_{G,lin}^{LI} = \kappa \left[ 2 \frh^{ab} \partial_{[c}
    C^{c} {}_{b]a} + 2 \eta^{ab} C^c {}_{a[b} C^d
    {}_{d]c} \right]
\end{equation}
with $\kappa \equiv 1/(16 \pi G)$, and
$\mathcal{L}_\mathrm{mat} [ \eta, \Psi^A ]$ is the Lagrange density
governing the ``matter sector'' of the action.  The matter sector is
assumed to depend on the fiducial metric density $\eta^{ab}$ as well
as some collection of matter fields $\Psi^A$.  Note that for
consistency in what follows, we will need to view $\eta^{ab}$ as a
tensor density rather than as a simple tensor.

The equations of motion derived from \eref{lag.LI.lin.Pal} are then
\begin{equation} \label{lin.EOM.Pal1}
  \partial_c C^c {}_{ab} - \partial_{(a} C^c {}_{b)c} = 0
\end{equation}
and
\begin{equation} \label{lin.EOM.Pal2}
  \partial_c \frh^{ab} - \partial_d \frh^{d(a} \delta^{b)} {}_c =
  \eta^{ab} C^d {}_{dc} + \eta^{de} C^{(a} {}_{de}
  \delta^{b)} {}_c - 2 \eta^{d(a} C^{b)} {}_{cd}.
\end{equation}
Some algebra can then show that \eref{lin.EOM.Pal2} is equivalent to
\begin{equation} \label{lin.EOM.Pal2b} C^c {}_{ab} = -
  \frac{1}{2} \left[ \eta_{da} \partial_b \bar{\frh}^{cd} + \eta_{db}
    \partial_a \bar{\frh}^{cd} - \eta_{ad} \eta_{be} \eta^{cf}
    \partial_f \bar{\frh}^{de} \right] \equiv \Gamma^c {}_{ab},
\end{equation}
where we have defined $\bar{\frh}^{ab} \equiv \frh^{ab} - \frac{1}{2}
\eta^{ab} \eta_{cd} \frh^{cd}$.  For future reference, the right-hand
side of this equation is defined to be $\Gamma^c {}_{ab}$;  the
notation is intentionally suggestive.

We now define a tensor density
$\frg^{ab} = \eta^{ab} + \frh^{ab}$.  This tensor density will be
related to an inverse metric $g^{ab}$ by
$\frg^{ab} = \sqrt{ -g} g^{ab}$, with $g \equiv \det(g_{\mu \nu})$
according to the flat spacetime coordinates.  This implies that
$\det (\frg_{\mu \nu}) = 1/\det(g_{\mu \nu})$, and so we have
\begin{equation}
  g^{ab} = \sqrt{ - \frg} \frg^{ab}.
\end{equation}
But if we define $h^{ab} = g^{ab} - \eta^{ab}$, then it is not hard to
show that to linear order in $\frh^{ab}$,
\begin{equation}
  h^{ab} \approx \bar{\frh}^{ab}.
\end{equation}
In other words, \eref{lin.EOM.Pal2b} implies that $C^c {}_{ab}$ is
the linearized Christoffel symbol $\Gamma^c {}_{ab}$ associated with
the inverse metric $g^{ab}$ and this coordinate basis.  Equation
\eref{lin.EOM.Pal1}, meanwhile, says that the linearized Ricci tensor
associated with this Christoffel symbol is zero.  Thus, at linear
order, the free gravitational action \eref{lag.LI.lin.Pal} yields
equations of motion that are equivalent to the linearized Einstein
equations.

We now wish to couple $\frh^{ab}$ to the stress-energy of the
model. As the left-hand side of \eref{lin.EOM.Pal1} is equal to the
Ricci tensor, we expect the source on the right-hand side to be the
trace-reversed stress-energy tensor $\tau_{ab}$.  This can be
accomplished by adding the term $-\frh^{ab} t_{ab}$ to the action
\eref{lag.LI.lin.Pal}.  The trace-reversed stress energy tensor
$\tau_{ab}$ can be found via the Rosenfeld prescription by
``promoting'' the fiducial metric density $\eta^{ab}$ to an auxiliary
metric density $\psi^{ab}$, differentiating the Lagrange density
$\mathcal{L}^{LI}_{(0)}$ with respect to this auxiliary density, and
then setting $\psi^{ab} \to \eta^{ab}$ in the result; this will yield
$- \frac{1}{2} \tau_{ab}$.  (In this process, factors of
$\sqrt{-\eta}$ may need to be inserted into the action to ensure that
various sums in our expressions have a definite weight.)  The
derivatives in the first term of \eref{lag.LI.lin.Pal} must also be
``promoted'' to covariant derivatives, which are then varied along
with $\psi^{ab}$.  The result is
\begin{equation} \label{lin.LI.SE}
  - \tau_{ab} = 2 \kappa \left[ C^{c} {}_{a[b} C^{d}
    {}_{d]c} +   \sigma_{ab} \right],
\end{equation}
where $\sigma_{ab}$ is a total derivative:\footnote{Note that the
  expression for this quantity in \cite{Kostelecky2009} contains a
  sign error in one term.}
\begin{eqnarray} \label{sigma.LI.def} \fl \sigma_{ab} = - \frac{1}{2}
  \partial_c \bigg[ 2 \frh^{cd} \eta_{d(a} C_{b)} - \eta_{ad}
  \eta_{be} \eta^{cf} \frh^{de} C_f + 2 \eta_{e(a} \eta_{b)f}
  \eta^{gc} \frh^{de} C^{f}{}_{dg} - 2 \eta_{e(a} \frh^{cd}
  C^e{}_{b)d} \nonumber \\ - 2 \eta_{e(a} \delta_{b)} {}^f \frh^{de} C^c {}_{df}
  - \eta_{ab} \eta_{df} \eta^{cg} \frh^{de} C^f {}_{eg} + \frac{1}{2}
  \eta_{ab} \eta_{de} \eta^{cf} \frh^{de} C_f \bigg],
\end{eqnarray}
where we have defined $C_a \equiv C^b {}_{ba}$.  I have explicitly
written out the fiducial metrics used to raise and lower indices in
this expression in order to illustrate a point that will arise later
in the Lorentz-violating case.

In Deser's procedure, the gravitational sector of the action is
completed by adding the \emph{non-derivative} portion of
\eref{lin.LI.SE} to the action, coupled to $\frh^{ab}$:
\begin{equation} \label{lag.LI.NL.Pal} \mathcal{L}_G^{LI} = \kappa
  \left[ 2 \frh^{ab} \partial_{[c} C^{c} {}_{b]a} + 2 (\eta^{ab}
    + \frh^{ab}) C^c {}_{a[b} C^d {}_{d]c} \right] 
\end{equation}
Importantly, these new terms do not refer to the fiducial metric
density $\eta^{ab}$ in any way; thus, this additional term will not
contribute to $\tau_{ab}$.  This is the signal advantage of Deser's
choice to use the tensor density as the fundamental field;  it does
not require a further infinite series of terms in the gravitational
action, as is necessary when viewing the metric perturbation $h_{ab}$
as the fundamental field.

If we define $\frg^{ab} = \eta^{ab} + \frh^{ab}$, the resulting
equations of motion from this non-linear action can then be written as
\begin{equation} \label{NL.EOM.Pal1}
    \partial_c C^c {}_{ab} - \partial_{(a} C^c {}_{b)c} +
    C^c {}_{ab} C^d {}_{dc} - C^c {}_{ac} C^{d}
    {}_{bd} = 0
\end{equation}
and
\begin{equation} \label{NL.EOM.Pal2} \partial_c \frg^{ab} - \partial_d
  \frg^{d(a} \delta^{b)} {}_c = \frg^{ab} C^d {}_{dc} + \frg^{de}
  C^{(a} {}_{de} \delta^{b)} {}_c - 2 \frg^{d(a} C^{b)}
  {}_{cd}.
\end{equation}
By taking various traces and linear combinations of this second
equation, it can be shown to be equivalent to
\begin{eqnarray} \fl C^c {}_{ab} = - \frac{1}{2} \Bigg[ 2 \frg_{d(a}
  \partial_{b)} \frg^{cd} - \frg_{ad} \frg_{be} \frg^{cf} \partial_f
  \frg^{de} \nonumber \\ {} + \frg_{de} \left( - \delta^c {}_{(a}
    \partial_{b)} \frg^{de} + \frac{1}{2} \frg_{ab} \frg^{cf}
    \partial_f \frg^{de}\right) \Bigg], \label{NL.EOM.Pal2b}
\end{eqnarray}
where $\frg_{ab}$ is the inverse of $\frg^{ab}$ (i.e., $\frg_{ab}
\frg^{bc} = \delta_a {}^c$.)  In other words, $C^c {}_{ab}$ is
equal to the Christoffel symbol associated with the metric $g^{ab} =
\sqrt{-\frg} \frg^{ab}$ (with $\frg \equiv \det(\frg_{ab})$), and in
the absence of other matter, the Ricci tensor \eref{NL.EOM.Pal1}
associated with these Christoffel symbols vanishes.

On the other hand, it turns out that \eref{NL.EOM.Pal1} and
\eref{NL.EOM.Pal2} are equivalent to coupling \eref{lin.EOM.Pal1} to
the \emph{full} stress-energy of the action \eref{lin.LI.SE},
including the derivative term $\sigma_{ab}$.  To show this, rewrite
\eref{NL.EOM.Pal2} as
\begin{eqnarray} \fl \label{NL.EOM.Pal2c} \eta^{ab} C^d {}_{dc} +
  \eta^{de} C^{(a} {}_{de} \delta^{b)} {}_c - 2 \eta^{d(a} C^{b)}
  {}_{cd} \nonumber \\ = \partial_c \frh^{ab} - \partial_d \frh^{d(a}
  \delta^{b)} {}_c - \left[ \frh^{ab} C^d {}_{dc} + \frh^{de} C^{(a}
    {}_{de} \delta^{b)} {}_c - 2 \frh^{d(a} C^{b)} {}_{cd} \right].
\end{eqnarray}
(Note that $\partial_a \frh^{bc} = \partial_a \frg^{bc}$.)  Via the
same algebraic procedure used to obtain \eref{NL.EOM.Pal2b} from
\eref{NL.EOM.Pal2}, we find that
\begin{eqnarray} \fl C^c {}_{ab} = - \frac{1}{2}
  \left[ 2 \eta_{d(a} \partial_{b)} \bar{\frh}^{cd} - 
    \eta_{ad} \eta_{be} \eta^{cf} \partial_f \bar{\frh}^{de} \right] \nonumber
  \\ + \frac{1}{2} \left[ 2 \eta_{d(a} \frh^{cd} C_{b)} - 2 \eta_{e(a}
    \frh^{de} C^c {}_{b)d} - 2 \eta_{e(b} \frh^{cd} C^e {}_{c)d} -
    \eta_{ae} \eta_{bf} \eta^{cd} \frh^{ef} C_d \right.  \nonumber \\
  \left. {} +2 \eta_{e(a} \eta_{b)f} \eta^{cg} \frh^{de} C^f {}_{dg} -
    \eta_{de} \delta^c {}_{(a} C_{b)} \frh^{de} + 2\eta_{df} \delta^c
    {}_{(a} \frh^{de} C^f {}_{b)e} \right.  \nonumber \\ \left. {} +
    \frac{1}{2} \eta_{ab} \eta^{cf} \eta_{de} C_f \frh^{de} -
    \eta_{ab} \eta^{cg} \eta_{df} \frh^{de} C^f {}_{eg}
  \right] \label{NL.EOM.Pal2d} 
\end{eqnarray}
where we have defined $C_a \equiv C^b {}_{ab}$.  The first set of
terms can be seen to be $\Gamma^c {}_{ab}$.  Taking the appropriate
derivatives and contractions of \eref{NL.EOM.Pal2d}, and after a fair
amount of algebra, we find that \eref{NL.EOM.Pal2} implies that
\begin{equation}
  \partial_c C^c {}_{ab} - \partial_{(a} C^c {}_{b)c} =
  \partial_c \Gamma^c {}_{ab} - \partial_{(a} \Gamma^c
  {}_{b)c} + \sigma_{ab}, 
\end{equation}
with $\sigma_{ab}$ (remarkably) defined as in \eref{sigma.LI.def}.
Combining this with \eref{NL.EOM.Pal1}, we obtain
\begin{equation} \label{lin.EOM.source}
\partial_c \Gamma^c {}_{ab} - \partial_{(a} \Gamma^c
  {}_{b)c} = - \left[ C^c {}_{ab} C^d {}_{dc} - C^c
    {}_{ad} C^{d} {}_{bc} + \sigma_{ab} \right] = \frac{1}{2
    \kappa} \tau_{ab},
\end{equation}
with $\tau_{ab}$ defined as in \eref{lin.LI.SE}. In other words, the
non-linear equations of motion \eref{NL.EOM.Pal1} and
\eref{NL.EOM.Pal2} are equivalent to the equations \eref{lin.EOM.Pal1}
and \eref{lin.EOM.Pal2} from the linear action \eref{lag.LI.lin.Pal},
with the full stress-energy of the linear action acting as a source.
Note that even though the derivative portion of the stress-energy
$\sigma_{ab}$ is not explicitly coupled to the density $\frh^{ab}$ in
the non-linear action \eref{lag.LI.NL.Pal}, the terms corresponding to
it still arise in \eref{lin.EOM.source} so long as the full non-linear
equations of motion \eref{NL.EOM.Pal1} and \eref{NL.EOM.Pal2} hold.  I
will return to this point when we pass to the Lorentz-violating
version of the theory.

To include the effects of matter, one must also apply the bootstrap
procedure to the matter Lagrange density $\mathcal{L}_\mathrm{mat}$.  So
long as the matter action $\mathcal{L}_\mathrm{mat} [\eta^{ab}, \Psi^A]$
only depends on the fiducial metric density $\eta^{ab}$ itself, and
not on its derivatives, it can be shown \cite{Kostelecky2009} that the
net effect of applying the bootstrap procedure to the matter action is
simply to replace $\eta^{ab}$ with $\frg^{ab}$.\footnote{In the Deser
  procedure, this process \emph{does} sometimes require an infinite
  series of terms that might not be required if we view $h_{ab}$ as
  the fundamental field.  One cannot always escape both Scylla and
  Charybdis.}  Any matter action only containing Lorentz scalars will
satisfy this condition, as well as any $n$-form field whose kinetic
terms depend only on that field's exterior derivative.  In particular,
the Maxwell kinetic term $- \frac{1}{4} F_{ab} F^{ab}$, with
$F_{ab} = 2 \partial_{[a} A_{b]}$, is independent of the choice of
derivative operator and so only depends on the metric itself and not
on the metric derivatives.  The trace-reversed stress-energy will then
appear on the right-hand side of \eref{NL.EOM.Pal1}, while
\eref{NL.EOM.Pal2} will be unaffected.

\subsection{Deser bootstrap procedure for Lorentz-violating
  gravity \label{LV.Deser.sec}}

In Section \ref{propagator.sec}, it was shown that under certain
assumptions, the only possible modification of the linearized Einstein
equation which couples the linearized metric perturbations to a
background vector $A_a$ is equivalent to replacing the fiducial metric
$\eta^{ab}$ with an effective metric:
\begin{equation} \label{effmetricdef2}
\tilde{\eta}^{ab} = \eta^{ab} + \xi \eta^{ac} \eta^{bd} A_c A_d.
\end{equation}
Since we are starting off from the context of flat spacetime, we will
assume that $A_a$ is a constant background vector field; only later
will we ascribe dynamics to it.

We now wish to do two things.  First, we wish to modify the linear
action \eref{lag.LI.lin.Pal} so that its equations of motion are
equivalent to \eref{lin.EOM}, with $\mathcal{P}^{abcdef}$ given by
\eref{genprop} and $\mathcal{Q}^{abcd} = \mathcal{R}^{abcef} = 0$.
Second, we want to self-consistently couple the stress-energy of the
resulting action to itself, to obtain a non-linear model of
Lorentz-violating gravity.

There are two possible terms we can add to the action
\eref{lag.LI.lin.Pal} to couple the fields to a constant background
vector field $A_a$:
\begin{eqnarray} \fl \mathcal{L}_{G,lin}^{LV} =
  \kappa \left[ 2 (\frh^{ab} + \xi_1 \eta^{ae} \eta^{bf} A_e A_f)
    \partial_{[c} C^{c} {}_{b]a} \right. \nonumber \\ \left. {} + 2
    (\eta^{ab}+ \xi_2 \eta^{ae} \eta^{bf} A_e A_f) C^c {}_{a[b}
    C^d {}_{d]c} \right], \label{lag.LV.lin.Pal}
\end{eqnarray}
where $\xi_1$ and $\xi_2$ are arbitrary coupling constants.  The first
term, with coupling constant $\xi_1$, does not affect the linear
equations of motion at all; since $A_a$ is a constant vector field,
this term is a total derivative.  It does, however, change the
stress-energy of the model, and it will become important for our
interpretation of the non-linear model.  The second term, with
coupling constant $\xi_2$, basically replaces the fiducial metric
density with the ``effective metric density'' given by
\eref{effmetricdef2} under the substitution $\xi \to \xi_2$.  At this
point, there is no particular relationship between $\xi_1$ and
$\xi_2$; however, we will find that the interpretation of the model is
much more compelling and elegant when they are equal.

In the interests of compactness, I will need to define various
versions of tensors and tensor densities that depend on $\xi_1$ and
$\xi_2$.  I will use $[i]$ as a prepended superscript to
denote a version of a quantity defined in Section \ref{LI.Deser.sec}
(or subsequently) under the substitution $\xi \to \xi_i$.  For
example, the addition of the $\xi_2$ term in \eref{lag.LV.lin.Pal}
effectively replaces
\begin{equation}
  \eta^{ab} \to {}^{[2]}\tilde{\eta}^{ab} = \eta^{ab} + \xi_2
  \eta^{ac} \eta^{bd} A_b A_d.
\end{equation}
For convenience in what follows, I will also define
\begin{equation}
  {}^{[i]}\tilde{\frh}^{ab} = \frh^{ab} + \xi_i \eta^{ae} \eta^{bd}
  A_b A_d.
\end{equation}

It is not hard to see that under these modifications, the linear
equations of motion for $C^c {}_{ab}$ and $\frh^{ab}$ are simply
\eref{lin.EOM.Pal1}, unchanged, and \eref{lin.EOM.Pal2} under the
substitution $\eta^{ab} \to {}^{[2]}\tilde{\eta}^{ab}$.  Thus, the
linear Lagrange density \eref{lag.LV.lin.Pal} yields the desired
Lorentz-violating linearized Einstein equation given by \eref{lin.EOM}
and \eref{genprop}, with the substitution $\xi \to \xi_2$.

We now wish to apply the bootstrap procedure to the action
\eref{lag.LV.lin.Pal}.  The stress-energy in \eref{lin.LI.SE} will
become
\begin{equation}
  \label{lin.LV.SE}
  - \tau_{ab} = 2 \kappa \left[ C^{c} {}_{a[b} C^{d}
    {}_{d]c} + {}^{[1]}\sigma_{ab} \right] - \kappa (\tau_\xi)_{ab},
\end{equation}
where $(\tau_\xi)_{ab}$ is the contribution to the stress-energy
tensor coming from the \emph{algebraic} appearances of $\eta^{ab}$ in
the new coupling terms in \eref{lag.LV.lin.Pal}, and
${}^{[1]} \sigma_{ab}$ is obtained from \eref{sigma.LI.def} under the
substitution $\frh^{ab} \to {}^{[1]} \tilde{\frh}^{ab}$.

In parallel with Deser's procedure, the method will be to again couple
the non-derivative portion of the stress-energy \eref{lin.LV.SE} to
the field $\frh^{ab}$, and to see whether these equations are
equivalent to the linear equations \eref{lin.EOM.Pal1}, with $-
\frac{1}{2} \tau_{ab}$ included as a source, and \eref{lin.EOM.Pal2} ,
with $\eta^{ab} \to {}^{[2]}\tilde{\eta}^{ab}$.  The $(\tau_\xi)_{ab}$
piece of the new stress-energy tensor can be viewed as new
contribution from the ``matter sector''.  It depends on the metric
$\eta^{ab}$ itself, and so we will need to apply the procedure to this
new term, adding in the non-derivative portion of the stress-energy
from this term.  Iterating this procedure, this will generate an
infinite sum of terms as we add the higher-order contributions of
these terms to the stress-energy.  The procedure is analogous to the
infinite series of terms that arises from the matter sector in the
Lorentz-invariant bootstrap procedure; and as in that case, the
resulting terms can be resummed, with the net effect of replacing
\begin{equation}
  \xi_i \eta^{ae} \eta^{bf} A_e A_f \to \xi_i \sqrt{ - \frg}{ \frg^{ae}
    \frg^{bf}} A_e A_f
\end{equation}
in \eref{lag.LV.lin.Pal}.  Including this modification, along with
the coupling between $\frh^{ab}$ and the first term of
\eref{lin.LV.SE}, the full non-linear action becomes
\begin{equation} \label{lag.LV.NL.Pal} \mathcal{L}_{G}^{LV} = 2 \kappa
  \left[ ({}^{[1]}\tilde{\frg}^{ab} - \eta^{ab}) \partial_{[c} C^{c}
    {}_{b]a} + {}^{[2]}\tilde{\frg}^{ab} C^c {}_{a[b}
    C^d {}_{d]c} \right],
\end{equation}
where we have defined
\begin{equation} {}^{[i]}\tilde{\frg}^{ab}
  = \frg^{ab} + \xi_i \sqrt{-\frg} \frg^{ae} \frg^{bf} A_e A_f.
\end{equation}
Note that in contrast with the Lorentz-invariant definition,
\begin{equation}
  {}^{[i]}\tilde{\frg}^{ab} \neq \eta^{ab} +
  {}^{[i]}\tilde{\frh}^{ab} = \eta^{ab} + \frh^{ab} + \xi_i \eta^{ae} \eta^{bd}
  A_b A_d.
\end{equation}
This difference will become important in what follows.

The equations of motion derived from \eref{lag.LV.NL.Pal} can be
obtained by viewing $\frg^{ab}$ and $C^c {}_{ab}$ as the configuration
variables.  In performing the variation with respect to $\frg^{ab}$,
it is useful to note that
\begin{eqnarray} 
  \frac{ \delta \left({}^{[i]}\frg^{cd}\right)}{\delta \frg^{ab}} &=
  \delta^c {}_{(a} \delta^d {}_{b)} + \xi_i \sqrt{-\frg} \left[ 2
    A_{(a} \delta_{b)} {}^{(c} \frg^{d)e} A_e - \frac{1}{2} \frg_{ab}
    \frg^{ce} \frg^{df} A_e A_f \right] \nonumber \\
  &=\delta^c {}_{(a} \delta^d {}_{b)} + \xi_i \left[ 2 A_{(a}
    \delta_{b)} {}^{(c} g^{d)e} A_e - \frac{1}{2} g_{ab}
    g^{ce} g^{df} A_e A_f \right],
\end{eqnarray}
and so the equation of motion associated with $\frg^{ab}$ becomes
\begin{eqnarray} \fl \partial_{c} C^c {}_{ab} - \partial_{(a} C_{b)} +
  2 C^c {}_{a[b} C^d {}_{d]c} \nonumber \\ {} + \xi_1 \left[ 2 A_{(a}
    \delta_{b)} {}^{(c} A^{d)} - \frac{1}{2} g_{ab} A^c A^d \right]
  \left( \partial_{e} C^e {}_{cd} - \partial_{(c} C_{d)} \right)
  \nonumber \\ + 2 \xi_2 \left[ 2 A_{(a} \delta_{b)} {}^{(c} A^{d)} -
    \frac{1}{2} g_{ab} A^c A^d \right] C^e {}_{c[d} C^f {}_{f]c} =
  \frac{1}{2 \kappa} (\tau_\mathrm{mat})_{ab}, \label{gen.LV.EOM.Pal1}
\end{eqnarray}
where all indices are raised and lowered with $g_{ab}$ and its
inverse.

Meanwhile, the only change for the $C^c {}_{ab}$ equation of motion,
relative to the Lorentz-invariant version \eref{NL.EOM.Pal2}, is that
we must substitute $\frg^{ab} \to {}^{[i]}\tilde{\frg}^{ab}$
appropriately:
\begin{equation} \label{gen.LV.EOM.Pal2}
  \partial_c {}^{[1]}\tilde{\frg}^{ab} - \partial_d
  {}^{[1]}\tilde{\frg}^{d(a} \delta^{b)} {}_c = {}^{[2]}\tilde{\frg}^{ab} C^d
  {}_{dc} + {}^{[2]}\tilde{\frg}^{de} 
  C^{(a} {}_{de} \delta^{b)} {}_c - 2 {}^{[2]}\tilde{\frg}^{d(a} C^{b)}
  {}_{cd}.
\end{equation}

I have been unable to find an elegant interpretation of the equations
of motion \eref{gen.LV.EOM.Pal1} and \eref{gen.LV.EOM.Pal2} in the
general $\xi_1 \neq \xi_2$ case.  Similar to the Lorentz-invariant
case, \eref{gen.LV.EOM.Pal2} can still be inverted to obtain an
expression for $C^c {}_{ab}$:
\begin{eqnarray} \fl C^c {}_{ab} = - \frac{1}{2} \left[ 2 {}^{[2]}
    \tilde{\frg}_{d(a} \partial_{b)} {}^{[1]} \tilde{\frg}^{cd} -
    {}^{[2]} \tilde{\frg}_{ad} {}^{[2]} \tilde{\frg}_{be} {}^{[2]}
    \tilde{\frg}^{cf} \partial_f {}^{[1]} \tilde{\frg}^{de}
  \right. \nonumber \\ \left. + {}^{[2]} \tilde{\frg}_{de} \left( -
      \delta^c {}_{(a} \partial_{b)} {}^{[1]} \tilde{\frg}^{de} +
      \frac{1}{2} {}^{[2]} \tilde{\frg}_{ab} {}^{[2]}
      \tilde{\frg}^{cf} \partial_f {}^{[1]} \tilde{\frg}^{de}\right)
  \right],\label{gen.LV.EOM.Pal2b}
\end{eqnarray}
where ${}^{[2]} \tilde{\frg}_{ab}$ is the inverse of
${}^{[2]} \tilde{\frg}^{ab}$.  If $\xi_1 \neq \xi_2$, $C^c {}_{ab}$
can no longer be interpreted as the Christoffel symbol associated with
either of the metrics ${}^{[1]} \tilde{g}_{ab}$ or
${}^{[2]} \tilde{g}_{ab}$.  Even if $C^c {}_{ab}$ could be interpreted
as the Christoffel symbol for some third metric, the fact that
$\xi_1 \neq \xi_2$ in \eref{gen.LV.EOM.Pal1} prevents us from
interpreting that equation in terms of the curvature of that metric.

However, if $\xi_1 = \xi_2 \equiv \xi$, then neither of these problems
arise.  In this case, $C^c {}_{ab}$ is simply the Christoffel symbol
associated with the inverse metric density
\begin{equation} \label{sym.unphys.density.def}
  \tilde{\frg}^{ab} = \tilde{\frg}^{ab}
  = \frg^{ab} + \xi \sqrt{-\frg} \frg^{ae} \frg^{bf} A_e A_f,
\end{equation}
and \eref{gen.LV.EOM.Pal1} simply becomes
\begin{eqnarray} \fl \left[ \delta^c {}_{(a}
    \delta^d {}_{b)} + \xi \left( 2 A_{(a} \delta_{b)} {}^{(c} g^{d)e}
      A_e - \frac{1}{2} g_{ab} g^{ce} g^{df} A_e A_f \right) \right]
  \left( \partial_{e} C^e {}_{cd} - \partial_{(c} C_{d)} + 2 C^e
    {}_{c[d} C^f {}_{f]c} \right) \nonumber \\ = \frac{1}{2 \kappa}
  (\tau_\mathrm{mat})_{ab}. \label{sym.LV.EOM.Pal1} 
\end{eqnarray}
The second factor on the left-hand side of \eref{sym.LV.EOM.Pal1} is
then equal to $\tilde{R}_{cd}$, the Ricci tensor of the gravitational
metric $\tilde{g}_{ab}$ given implicitly by the relationship
$\tilde{\frg}^{ab} = \sqrt{-\tilde{g}} \tilde{g}^{ab}$.  The equations
of motion from the action \eref{lag.LV.NL.Pal} (with $\xi_1 = \xi_2 =
\xi$) are therefore equivalent to
\begin{equation} \label{full.LV.EOM.secorder}
 \tilde{R}_{ab} + 2 \xi A_{(a} \tilde{R}_{b)c} A^c - \frac{1}{2} \xi
 g_{ab} \tilde{R}_{cd} A^c A^d = 8 \pi G (\tau_\mathrm{mat})_{ab},
\end{equation}
where all indices are raised and lowered with $g^{ab}$ and its
inverse.  Given the relative straightforwardness of this particular
case, I will be assuming that $\xi_1 = \xi_2 = \xi$ for the remainder
of this work.

It can be shown that if $\tilde{\frg}^{ab}$ and $\frg^{ab}$ are
related by \eref{sym.unphys.density.def}, then the corresponding
undensitized inverse metrics are related by
\begin{equation} \label{inverse.met.def}
  \tilde{g}^{ab} = \frac{1}{\sqrt{1 + \xi A^2}} \left( g^{ab} + \xi
    g^{ae} g^{bf} A_e A_f \right),
\end{equation}
and the metrics themselves are related by
\begin{equation} \label{met.def}
  \tilde{g}_{ab} = \sqrt{1 + \xi A^2} g_{ab} - \frac{\xi}{\sqrt{1 +
      \xi A^2} }A_a A_b
\end{equation}
with $A^2 \equiv A_a A_b g^{ab}$.  The determinants of these metrics,
meanwhile, are related by
\begin{equation} \label{det.def}
  \det (\tilde{g}_{\mu \nu}) = (1 + \xi A^2) \det(g_{\mu \nu}).
\end{equation}

This action can be viewed as a natural generalization of Deser's
bootstrap procedure to a Lorentz-violating gravity model.  However, the
interpretation of this procedure as ``coupling $\frh^{ab}$ to its own
stress-energy'', a particularly elegant feature of the
Lorentz-invariant model, does not carry over nicely to the present
case.  (I will ignore the matter sector for the remainder of this
section, as it does not affect the following argument.)  Recall that
in the Lorentz-invariant case, we added a coupling between $\frh^{ab}$
and the \emph{non-derivative} part of the stress-energy tensor.  We
were then able to interpret the resulting non-linear equations of
motion \eref{NL.EOM.Pal1} and \eref{NL.EOM.Pal2} in terms of the the
linear equations of motion coupled to the \emph{full} stress-energy
tensor of the linear model; the derivation from Equations
\eref{NL.EOM.Pal2c} to \eref{lin.EOM.source} showed how the
derivative portion of the stress-energy tensor $\sigma_{ab}$ emerged
from the non-linear model naturally.

The presence of the Lorentz-violating terms disrupts this line of
logic severely.  In the case where $\xi_1 = \xi_2 = \xi \neq 0$, the
derivative portion of the stress-energy is now
\begin{eqnarray} \fl \sigma_{ab} = - \frac{1}{2} \partial_c \bigg[ 2
  \tilde{\frh}^{cd} \eta_{d(a} C_{b)} - \eta_{ad} \eta_{be} \eta^{cf}
  \tilde{\frh}^{de} C_f + 2 \eta_{e(a} \eta_{b)f} \eta^{gc}
  \tilde{\frh}^{de} C^{f}{}_{dg} - 2 \eta_{e(a} \tilde{\frh}^{cd}
  C^e{}_{b)d} \nonumber \\ - 2 \eta_{e(a} \delta_{b)} {}^f
  \tilde{\frh}^{de} C^c {}_{df} - \eta_{ab} \eta_{df} \eta^{cg}
  \tilde{\frh}^{de} C^f {}_{eg} + \frac{1}{2} \eta_{ab} \eta_{de}
  \eta^{cf} \tilde{\frh}^{de} C_f \bigg], \label{sigma.LV.def}
\end{eqnarray}
and the analogue of \eref{NL.EOM.Pal2c} becomes
\begin{eqnarray} \fl \tilde{\eta}^{ab} C^d {}_{dc} + \tilde{\eta}^{de}
  C^{(a} {}_{de} \delta^{b)} {}_c - 2 \tilde{\eta}^{d(a} C^{b)}
  {}_{cd} \nonumber \\ = \left( \partial_c \frh^{ab} - \partial_d
    \frh^{d(a} \delta^{b)} {}_c\right) - \left( \frh^{ab} C^d
    {}_{dc} + \frh^{de} C^{(a} {}_{de} \delta^{b)} {}_c - 2
    \frh^{d(a} C^{b)} {}_{cd} \right)
  \nonumber \\
  + \left( \partial_c \frk^{ab} - \partial_d \frk^{d(a} \delta^{b)}
  {}_c\right) - \left( \frl^{ab} C^d {}_{dc} + \frl^{de} C^{(a}
  {}_{de} \delta^{b)} {}_c - 2 \frl^{d(a} C^{b)} {}_{cd}
  \right) \label{sym.LV.NL.EOM.Pal2c} 
\end{eqnarray}
where I have defined
\begin{equation}
  \frk^{ab} = \xi \sqrt{-\frg} \frg^{ac} \frg^{bd}
\end{equation}
and
\begin{equation}
  \frl^{ab} = \xi \left[ \sqrt{-\frg} \frg^{ac} \frg^{bd} - 2
    \eta^{ac} \eta^{bd} \right] A_c A_d. 
\end{equation}
It does not seem possible to parallel the Lorentz-invariant derivation
any further from this point.  The next step would be to isolate $C^c
{}_{ab}$ from \eref{sym.LV.NL.EOM.Pal2c}, to obtain an analog of
\eref{NL.EOM.Pal2d}.  However, this would lead to terms involving
$\frk^{ab}$ and $\frl^{ab}$ (arising from the third and fourth sets of
terms on the right-hand side of \eref{sym.LV.NL.EOM.Pal2c},
respectively) that do not have an analog in the linear equations of
motion.  Moreover, the process of isolating $C^c {}_{ab}$ from
\eref{sym.LV.NL.EOM.Pal2c} in the present case would involve raising
and lowering indices with the metric $\tilde{\eta}^{ab}$, rather than
the fiducial metric $\eta^{ab}$.  The terms that enter into the
derivative portion of the stress-energy $\sigma_{ab}$, however, have
their indices raised and lowered with $\eta^{ab}$.  There do not
appear to be any fortuitous cancellations in all of these extra terms.
It seems that the equations of motion for the non-linear model
\eref{lag.LV.NL.Pal} cannot easily be interpreted as ``the linear
field $\frh^{ab}$ coupled to its own stress-energy.''

The presence of a fixed background vector field $A_a$ on the
underlying flat spacetime means that the equations of motion will not
have diffeomorphism invariance; the fixed background vector field
explicit breaks this symmetry.  While some recent work has explored
the possibilities of Lorentz symmetry violation via an explicitly
breaking of diffeomorphism invariance \cite{Bluhm2017}, a more common
tactic is to restore diffeomorphism invariance to these equations by
``promoting'' the background vector field $A_a$ to a dynamical field
\cite{Kostelecky2004}.  The simplest way to do this is by assuming
that $A_a$ is governed by the action
\begin{equation} \label{LV.field.action.flat}
  \mathcal{L}_A = - \frac{1}{4} \eta^{ac} \eta^{bd} F_{ab} F_{cd} - V(
  A_a A_b \eta^{ab})
\end{equation}
in flat spacetime, where $V(A_a A_b \eta^{ab})$ is a Higgs-like
potential energy for $A_a$ that is responsible for the breaking of
Lorentz symmetry.  When we apply the bootstrap procedure, this will
simply replace the inverse metric density $\eta^{ab}$ with $\frg^{ab}
= \eta^{ab} + \frh^{ab}$ throughout; the result will then be
\begin{eqnarray} \fl \mathcal{L}_A = - \frac{1}{4} \sqrt{-\frg}
  \frg^{ac} \frg^{bd} F_{ab} F_{cd} - \frac{1}{\sqrt{-\frg}}
  V\left(\sqrt{-\frg} \frg^{ab} A_a A_b \right)
  \nonumber \\
  = \sqrt{-g} \left[ - \frac{1}{4} g^{ac} g^{bd} F_{ab} F_{cd} - V
  (A_a A_b g^{ab}) \right]\label{LV.field.action}
\end{eqnarray}

\subsection{Formalisms and frames}

\subsubsection{Palatini \& metric formalisms}

All told, our bootstrapped Lorentz-violating action is
\begin{eqnarray} \fl \mathcal{L} = 2 \kappa \tilde{\frg}^{ab} \left[
    \partial_{[c} C^{c} {}_{b]a} + C^c {}_{a[b} C^d {}_{d]c} \right]
  \nonumber \\ - \frac{1}{4} \sqrt{-\frg} \frg^{ac} \frg^{bd} F_{ab}
  F_{cd} - \frac{1}{\sqrt{-\frg}} V\left(\sqrt{-\frg} \frg^{ab} A_a
    A_b \right) +
  \mathcal{L}_{\mathrm{mat}}. \label{full.LV.NL.action.Pal.dens}
\end{eqnarray}
In terms of the metrics $\tilde{g}^{ab}$ and $g^{ab}$, this becomes
\begin{eqnarray}
  \fl \mathcal{L} = 2 \kappa
  \sqrt{-\tilde{g}} \tilde{g}^{ab} \left[ \partial_{[c} C^{c} {}_{b]a}
  + C^c {}_{a[b} C^d {}_{d]c} \right] \nonumber \\ + \sqrt{-g} \left[ -
  \frac{1}{4} g^{ac} g^{bd} F_{ab} F_{cd} - V (A_a A_b g^{ab})
  \right] + \mathcal{L}_{\mathrm{mat}}. \label{full.LV.NL.action.Pal}
\end{eqnarray}

Thus far, we have effectively been using a ``Palatini'' (first-order)
form of the gravitational action; the resulting equations of motion
are given by \eref{full.LV.EOM.secorder}.  However, one can also
obtain the same equations of motion in a more familiar ``metric''
(second-order) formalism, where the connection is viewed as a function
of the metric rather than as an independent field:
\begin{equation}
  \label{full.LV.NL.action.metric} \mathcal{L} = \kappa
  \sqrt{-\tilde{g}} \tilde{g}^{ab} \tilde{R}_{ab} + \sqrt{-g} \left[ -
    \frac{1}{4} g^{ac} g^{bd} F_{ab} F_{cd} - V (A_a A_b g^{ab})
  \right] + \mathcal{L}_{\mathrm{mat}}.
\end{equation}
To see that these are equivalent, it suffices to vary the
gravitational portion of the action.  We can view $g^{ab}$ and $A_a$
as our fundamental fields, with variations $\delta g^{ab}$ and $\delta
A_a$.  The gravitational metric $\tilde{g}^{ab}$ will then change by
$\delta \tilde{g}^{ab}$ under these variations.  Thus, the variation
of the gravitational portion of the action
\eref{full.LV.NL.action.metric} is
\begin{eqnarray}
  \fl \int d^4x \delta( \sqrt{-\tilde{g}} \tilde{g}^{ab} \tilde{R}_{ab} )
  = \int d^4 x \left[ \delta( \sqrt{-\tilde{g}} \tilde{g}^{ab})
  \tilde{R}_{ab} + \sqrt{-\tilde{g}}
  \tilde{g}^{ab} \delta( \tilde{R}_{ab}) \right] \nonumber \\
  = \int d^4 x \left\{ \delta( \sqrt{-\tilde{g}} \tilde{g}^{ab})
  \tilde{R}_{ab} +\sqrt{-\tilde{g}} \tilde{\nabla}_a \left[ \left( -
  \tilde{g}^{ab} \tilde{g}^{cd} + \tilde{g}^{ad} \tilde{g}^{bc}
  \right) \tilde{\nabla}_b \delta \tilde{g}_{cd} \right]\right\},
\end{eqnarray}
where $\tilde{\nabla}_a$ is the covariant derivative defined by
$\tilde{\nabla}_a \tilde{g}_{bc} = 0$.  The second term can be seen to
be a total derivative, and so it will not contribute to the equations
of motion.  Moreover, from \eref{inverse.met.def} and \eref{det.def}
it can be seen that
\begin{equation} \label{useful.metric.relation}
  \sqrt{-\tilde{g}} \tilde{g}^{ab} = \sqrt{-g} \left( g^{ab} + \xi A^a
    A^b \right),
\end{equation}
where $A^a \equiv g^{ab} A_b$. Thus,
\begin{eqnarray} \fl \delta( \sqrt{-\tilde{g}} \tilde{g}^{ab})
  \tilde{R}_{ab} = \sqrt{-g} \left[ \delta g^{ab} + 2 \xi \left(
      \delta g^{ac} A^{bd} A_c + A^{a} g^{bd} \delta A_d\right)
  \right.  \nonumber \\ \left. {} - \frac{1}{2} g_{cd} \delta g^{cd}
    \left( g^{ab} + \xi A^a A^b \right) \right] \tilde{R}_{ab}.
\end{eqnarray}
The equation of motion arising from the variation of the metric
$g^{ab}$ in this formalism is thus
\begin{equation}\label{full.LV.EOM.Jordan}
  \tilde{R}_{ab} - \frac{1}{2} g_{ab} g^{cd} \tilde{R}_{cd} + 2 \xi
  A_{(a} \tilde{R}_{b) c} A^c - \frac{1}{2} \xi g_{ab} A^c A^d
  \tilde{R}_{cd} = 8 \pi G T_{ab},
\end{equation}
where $T_{ab}$ here includes the stress-energy contributions of both
the vector field $A_a$ and any other matter sources present;  in
particular, we have for the vector field
\begin{equation} \label{vector.SE.Jordan}
  T_{ab} = F_a {}^c F_{bc} - \frac{1}{4} g_{ab} F_{cd} F^{cd} +
  g_{ab} V(A^2) + 2 A_a A_b V'(A^2),  
\end{equation}
where $A^2 \equiv g^{ab} A_a A_b$, and all indices have been raised
and lowered with $g^{ab}$.  Equation \eref{full.LV.EOM.Jordan} can be
seen to be equivalent to \eref{full.LV.EOM.secorder} by taking a
contraction of that equation with $g^{ab}$. (For a more general
discussion of when the Palatini and metric formalisms are equivalent,
see \cite{Iglesias2007,Lanczos1957}.)

The equation of motion arising from the variation of $A_a$ in
\eref{full.LV.NL.action.metric}, meanwhile, is
\begin{equation} \label{vector.EOM.Jordan}
  \nabla_b F^{ba} - V'(A^2) A^a + 2 \xi A^b \tilde{R}_b {}^a = 0,
\end{equation}
where all indices have been raised and lowered with $g^{ab}$.

Finally, it should be noted that the Ricci tensor associated with the
gravitational metric $\tilde{g}^{ab}$ is related to the Ricci tensor
of the fiducial metric by
\begin{equation} \label{Ricci.relation}
  \tilde{R}_{ab} = R_{ab} + \nabla_c \frC^c {}_{ab} - \nabla_a \frC^c
  {}_{cb} + \frC^c {}_{cd} \frC^d {}_{ab} - \frC^c {}_{da} \frC^d {}_{cb},
\end{equation}
where
\begin{equation} \label{Ctensor.def}
  \frC^c {}_{ab} \equiv \frac{1}{2} \tilde{g}^{cd} \left[ \nabla_a
    \tilde{g}_{bd} + \nabla_b \tilde{g}_{ad} - \nabla_d \tilde{g}_{ab}
  \right].  
\end{equation}
These relationships would allow the equations of motion
\eref{full.LV.EOM.Jordan} and \eref{vector.EOM.Jordan} to be fully
expressed in terms of the variables $g_{ab}$ and $A_a$.  However, the
relationships between $\tilde{g}_{ab}$ and $g_{ab}$, as given in
equations \eref{inverse.met.def} and \eref{met.def}, result in
expressions that are rather complicated, and so we will not exhibit
them explicitly here.

\subsubsection{Jordan \& Einstein frames}

In taking the equations of motion from the action
\eref{full.LV.NL.action.metric}, we can make a choice of which fields
we view as fundamental.  In particular, we can either choose to view
$g^{ab}$ or $\tilde{g}^{ab}$ as the fundamental metric in the theory,
and vary the action with respect to one or the other along with the
vector field $A_a$.  Since the relationship between the sets of field
variables $\{g^{ab},A_a\}$ and $\{\tilde{g}^{ab}, A_a\}$ is
invertible, we will obtain equivalent sets of equations of motion with
either choice. Using terminology from scalar-tensor gravity theories,
viewing $g^{ab}$ as the fundamental field (as was done in the previous
subsection) corresponds to working in the \emph{Jordan frame}, while
viewing $\tilde{g}^{ab}$ as the fundamental field corresponds to
working in the \emph{Einstein frame} \cite{Flanagan2004}.

When working in the Einstein frame, the variation of the gravitational
portion of the action is straightforward and familiar.  However, the
variation of the vector portion of the action (as well as any other
matter sources that might be present) is complicated by the fact that
they depend on the fiducial metric $g^{ab}$ rather than the
gravitational metric $\tilde{g}^{ab}$.  To derive the equations of
motion in the Einstein frame, we begin by contracting
\eref{inverse.met.def} with $A_a A_b$, yielding
\begin{equation} \label{Anormtilde.def}
  \tilde{A}^2 \equiv \tilde{g}^{ab} A_a A_b = A^2 \sqrt{ 1 + \xi A^2}.
\end{equation}
We can then invert \eref{inverse.met.def} and
\eref{met.def} to obtain
\begin{equation}
  g^{bc} = \sqrt{1 + \xi A^2} \tilde{g}^{bc} - \frac{\xi}{1 + \xi A^2}
  \tilde{g}^{be} \tilde{g}^{cf} A_e A_f
\end{equation}
Note that \eref{Anormtilde.def} could in principle be inverted to
yield a closed-form expression for $A^2$ in terms of $\tilde{A}^2$;
but this involves taking the root of a cubic polynomial, leading to
complicated expressions.  Instead, we can simply view $A^2$ as an
function of $\tilde{g}^{ab}$ and $A_a$, defined implicitly by
\eref{Anormtilde.def}.  In particular, varying both sides of
\eref{Anormtilde.def} with respect to $\tilde{g}^{ab}$ yields the
relation
\begin{equation} \label{Ntens.def} \mathcal{N}_{ab} = \frac{\delta
    (A^2)}{\delta \tilde{g}^{ab}} = \frac{\sqrt{1 + \xi A^2}}{1 +
    \frac{3}{2} \xi A^2} A_a A_b.
\end{equation}

With all of this in hand, we can write out the full equations of
motion for this model.  When we vary $A_a$, we obtain
\begin{eqnarray}
  \fl (\mathcal{E}_A)^a
  & \equiv \frac{1}{\sqrt{-\tilde{g}}} \frac{\delta
    \mathcal{L}}{\delta A_a} \nonumber \\
  &= \tilde{\nabla}_b \left( \sqrt{ \mathcal{Q}}
    g^{bc} g^{ad} F_{cd} 
    \right) - \sqrt{ \mathcal{Q}} \mathcal{M}^{abc} g^{de} F_{bd}
    F_{ce} \nonumber \\
  & \qquad {}- \frac{ g^{ab} A_b}{\mathcal{Q}^{3/2}} \left(  -
    \frac{\xi }{4} g^{ac} g^{bd} F_{ab} F_{cd} - \xi V(A^2) + 2
    \mathcal{Q} V'(A^2) \right), \label{full.NL.EOM.A}
\end{eqnarray}
where we have defined $\mathcal{Q} \equiv 1 + \xi A^2$ and
\begin{equation}
  \mathcal{M}^{abd} = \frac{1}{2} \frac{\delta g^{bc}}{\delta A_a} =
  \frac{\xi}{\mathcal{Q}} \left( - \tilde{A}^{(b} g^{c)a} +
    \frac{1}{\mathcal{Q}} \tilde{A}^b \tilde{A}^c g^{ad}A_d \right)
\end{equation}
with $\tilde{A}^a \equiv \tilde{g}^{ab} A_b$.  The equation of motion
obtained when we vary $\tilde{g}^{ab}$, meanwhile, is
\begin{eqnarray}
  \fl (\mathcal{E}_{\tilde{g}})_{ab}
  &\equiv \frac{1}{\sqrt{-\tilde{g}}}
    \frac{\delta \mathcal{L}}{\delta \tilde{g}^{ab}} \nonumber \\
  &= \kappa \tilde{G}_{ab} -
    \frac{1}{2} \sqrt{\mathcal{Q}} F_{ac} F_{bd} \tilde{g}^{cd} +
    \frac{\xi}{\mathcal{Q}} \left( A_{(a} F_{b)c} F_{de} \tilde{g}^{ce}
    \tilde{A}^d + \frac{1}{2} \tilde{A}^c \tilde{A}^d F_{c(a} F_{b)d}
    \right) \nonumber \\
  & \qquad {} - \frac{1}{2} \tilde{g}_{ab} \left( - \frac{1}{4} g^{ac}
    g^{bd} F_{ab} F_{cd} - 
    V(A^2)  \right) \nonumber \\
  & \qquad {} + \frac{1}{\sqrt{\mathcal{Q}}} \left( - \frac{\xi}{8}
    F_{ab} 
    \tilde{F}^{ab} - \frac{\xi}{2 \mathcal{Q}^{3/2}} \tilde{A}^a
    \tilde{A}^c \tilde{g}^{bd} F_{ab} F_{cd} + \frac{\xi}{2\mathcal{Q}}
    V(A^2) - V'(A^2) \right) \mathcal{N}_{ab}, \label{full.NL.EOM.tg}
\end{eqnarray}
with $\mathcal{N}_{ab}$ defined as in \eref{Ntens.def} and
$\tilde{F}^{ab} \equiv \tilde{g}^{ac} \tilde{g}^{bd} F_{bd}$.


\section{Applications \& connections \label{apply.sec}}

\subsection{SME coefficients}

The primary motivation of this work was to extend a Lorentz-violating
model of linearized gravity \cite{Bailey2006} to a fully non-linear
model.  Still, this model should still have a linearized limit, and
should be able to make predictions about the behavior of objects
moving under the influence of gravity in (for example) the solar
system.  Within the SME, a substantial machinery has been developed to
address such questions.  In the gravity sector \cite{Bailey2006}, the
observational effects will be parametrized by a set of ten
coefficients: a scalar $\bar{u}$ and a trace-free tensor
$\bar{s}^{ab}$. (If one allows higher-order equations of motion, the
situation is more complicated \cite{Kostelecky2016}.  However, since
we required in Section \ref{propagator.sec} that the equations of
motion only be second-order in derivatives of the metric, such effects
will not be present in this model.)

In general, to find the SME coefficients for a general gravitational
model, one must take the Euler-Lagrange equations, linearize them, and
combine them to yield an effective equation for the linearized Ricci
tensor.  In the process, one can typically only work to first order in
the parameter which controls Lorentz violation ($\xi$ in the present
case); there are often terms of order $\xi^2$ which are discarded in
the process, under the assumption that they will be negligible.  It is
therefore legitimate to expand \eref{full.LV.NL.action.metric} to
$\mathcal{O}(\xi)$, with the understanding that the effective SME
gravity equation will only be accurate to this order in any event.
More pragmatically, we will see that the action simplifies greatly in
this limit.

The Ricci tensor corresponding to $\tilde{g}^{ab}$ is given by
\eref{Ricci.relation} and \eref{Ctensor.def}.  We have, to
$\mathcal{O}(\xi)$,
\begin{equation}
  \tilde{g}_{ab} = g_{ab} + \xi \left( \frac{1}{2} A^2 g_{ab} - A_a
    A_b \right) + \mathcal{O}(\xi^2),
\end{equation}
and so
\begin{equation}
  \fl \frC^c {}_{ab} = \frac{\xi}{2} \left[ \delta^c {}_{(a} \nabla_{b)}
    \left( A^2 \right) - 2 \nabla_{(a} \left( A_{b)} A^c \right)
    - \frac{1}{2} g_{ab} \nabla^c \left( A^2 \right) +
    \nabla^c \left(A_a A_b \right) \right] + \mathcal{O}(\xi^2).
\end{equation}
In other words, $\frC^c {}_{ab}$ is $\mathcal{O}(\xi)$, and so the
gravitational portion of the action is
\begin{equation}
   \fl \sqrt{ - \tilde{g}} \tilde{g}^{ab} \tilde{R}_{ab} =
  \sqrt{-g}\left[ R + \xi A^a A^b R_{ab}  + \nabla_a \left( g^{bc}
      \frC^a {}_{bc} - g^{ab} \frC^c {}_{cb} 
    \right) \right] + \mathcal{O}(\xi^2). 
\end{equation}
The terms involving the derivatives of $\frC^c {}_{ab}$ are total
derivatives and so will not contribute to the equations of motion,
while the terms quadratic in $\frC^c {}_{ab}$ are higher-order in
$\xi$.

Since we can ignore these terms, and using the relation
\eref{useful.metric.relation} derived earlier, the non-linear action
is
\begin{equation}
  \mathcal{L} \approx \sqrt{-g}\left[ \kappa (R + \xi A^a A^b R_{ab})
    - \frac{1}{4} g^{ac} g^{bd} F_{ab} F_{cd} - 
    V(A^2) \right] 
\end{equation}
to this order in $\xi$.  This can be recognized as the action for the
so-called \emph{bumblebee model} \cite{Kostelecky2004}.  The SME
analysis for this equation was carried out in \cite{Bailey2006}, with
the results
\begin{equation}
  \bar{s}^{\mu \nu}
  = \xi \left( A^\mu A^\nu - \frac{1}{4} \eta^{\mu
      \nu} A^2 \right), \qquad  \bar{u} = -\frac{1}{12} \xi A^2.
\end{equation}
The bootstrapped Lorentz-violating model
\eref{full.LV.NL.action.metric} will therefore have these same SME
coefficients when we look at linearized solutions about a background
where the matter metric is Minkowski ($g^{ab} = \eta^{ab}$) and the
vector field $A_a$ is constant.

The components of the tensor $\bar{s}^{\mu \nu}$ in the Sun-centered Frame
\cite{Bluhm2003} have been bounded, directly or indirectly, by a
variety of experiments \cite{Kostelecky2008}.  For experiments within
the Solar System, the magnitudes of these components are $10^{-5}$
for $\bar{s}^{TT}$, $10^{-8}$ for $\bar{s}^{TI}$ ($I \in \{X, Y, Z\}$), and
$10^{-10}$ for $\bar{s}^{IJ}$; these bounds come from a combination of
precision gravimetry \cite{Shao2018} and lunar laser ranging
measurements \cite{Bourgoin2017}.  These constraints can be viewed as
bounding various combinations of the coupling constant $\xi$ and the
components of the vector field $A_\mu$ in the current cosmological
environment.

More stringent bounds on the components $\bar{s}^{\mu \nu}$, down to the
$10^{-14}$ level, have also been inferred from sources outside the
solar system, from observations of cosmic rays \cite{Kostelecky2015}.
While these latter bounds are indirect, requiring some assumptions
about the origin and nature of cosmic rays, they do imply that $\xi$
and/or the components $A^\mu$ must be quite small in the current
epoch.  The gravitational wave event GW170817, which was observed in
conjunction with a gamma-ray burst, also bounded certain combinations
of the $\bar{s}^{\mu \nu}$ components to the $10^{-15}$ level \cite{Abbott2017}.
While this bound is direct, it is worth noting that a single event
such as GW170817 only bounds the difference between the speeds of
electromagnetic and gravitational waves for one particular direction
of propagation, and thus only places bounds on a single combination of
the components $\bar{s}^{\mu \nu}$.  At the present time, the region of
parameter space consistent with these observations is still unbounded;
but as LIGO sees more such events, we would expect this region to
become rather stringently bounded.

\subsection{Generalized Proca theory}

The bootstrapped Lorentz-violating model
\eref{full.LV.NL.action.metric} can be connected to \emph{generalized
  Proca theory} \cite{Heisenberg2014, Heisenberg2017}.  Such models
were constructed as a Galileon-like generalization of Proca theory,
and generically include derivative self-interactions.  The general
form of the kinetic terms for the vector field in a generalized Proca
theory is
\begin{equation} \label{gen.Proca.form} \mathcal{L}_K = - \frac{1}{4}
  F_{ab} F^{ab} + \sum_{n = 2}^6 \beta_n \mathcal{L}_n,
\end{equation}
where the $\beta_i$'s are arbitrary coefficients, and $\mathcal{L}_2$ is
an arbitrary algebraic function $G_2$ of $F_{ab}$ and $A^a$.  The
terms $\mathcal{L}_i$ ($3 \leq i \leq 6$) depend on the
symmetric part $\nabla_{(a} A_{b)}$ of the derivative of the vector
field and on arbitrary algebraic functions $G_3$ through $G_6$.  The
precise form of these terms can be found in the above references;
however, we will see shortly that these terms vanish in the present
case.

To cast bootstrapped Lorentz-violating gravity into the above form, we
can rewrite \eref{full.LV.NL.action.metric} in the Einstein frame
using \eref{inverse.met.def} and \eref{det.def}.  The result is
\begin{eqnarray}
  \fl
  \mathcal{L} = \kappa
  \sqrt{-\tilde{g}} \left[ \tilde{g}^{ab} \tilde{R}_{ab} -
  \frac{1}{4} \sqrt{\mathcal{Q}} \tilde{g}^{ac} \tilde{g}^{bd} F_{ab}
  F_{cd} \right. \nonumber \\ \left. {} + \frac{1}{2} \frac{\xi}{\mathcal{Q}} \tilde{A}^a
  \tilde{A}^c \tilde{g}^{bd} F_{ab} F_{cd}  - V (A^2)
  \right] + \mathcal{L}_{\mathrm{mat}}[g].
\end{eqnarray}
The kinetic term for $A_a$ can clearly be seen to be a function of
$F_{ab}$ and $A^a$ only; in other words, we have
\begin{equation}
  \mathcal{L}_2 = - \frac{\sqrt{\mathcal{Q}} - 1}{4} \tilde{g}^{ac} \tilde{g}^{bd} F_{ab}
  F_{cd} + \frac{1}{2} \frac{\xi}{\mathcal{Q}} \tilde{A}^a
  \tilde{A}^c \tilde{g}^{bd} F_{ab} F_{cd},
\end{equation}
and the remaining terms in \eref{gen.Proca.form} vanish.  Since this
model is a special case of generalized Proca theory, this implies that
the bootstrapped Lorentz-violating model has the same desirable
properties as generalized Proca theory; in particular, it is free of
ghost instabilities and has only three propagating degrees of freedom
associated with the vector field.

Finally, I will make two related observations concerning this
connection. First, in most work involving generalized Proca theory, it
is assumed that the ``matter sector'' couples minimally to the
gravitational metric.  This is not the case here; to put the action in
the form \eref{gen.Proca.form}, it was necessary to work in the
Einstein frame.  This means that the cosmological solutions found in
\cite{Heisenberg2016} (and similar works) would not necessarily be
solutions of the current model, since the matter couples to the
gravitational fields differently.  This is an example of an
observation made (and exploited) in a recent work by
G\"umr\"uk\c{c}\"uo\v{g}lu and Koyama \cite{Gumrukcuoglu2019}: while
it is possible to find equivalent descriptions of a ``pure gravity''
action in terms of different frames, the choice of coupling between
``conventional matter'' and the gravitational fields can break this
equivalency.

Second, consider again this model in the Jordan frame.  While the
action \eref{full.LV.NL.action.metric} would be rather complicated if
written entirely in terms of $g_{ab}$, $\nabla_a$, and $A_a$, it is
obvious that it would not fit naturally into the class of theories
described in \cite{Heisenberg2014, Heisenberg2017}.  In particular,
the resulting action in our current model would contain a term of the
form $A^a A^b R_{ab}$, which is not included in any of the
$\mathcal{L}_i$ terms in the generalized Proca Lagrangian.  But we
know that (at least in the absence of matter) this model is equivalent
to a special case of generalized Proca theory.  This suggests that
there may be more models having the desirable properties of the
``original'' generalized Proca theories \cite{Heisenberg2014,
  Heisenberg2017} that have not yet been described.  Such models could
be obtained using similar techniques to those described in
\cite{Gumrukcuoglu2019}:  a change of frame in the gravitational
sector, followed by a minimal coupling between the ``new'' metric and
conventional matter.

\subsection{FRW cosmology}

As a simple illustration of a non-linear solution of this model, we
ask what a dark-energy-dominated FRW spacetime would look like in this
case.  If we assume that our solution is spatially homogenous and
isotropic, the form of the gravitational metric must be of the
standard FRW form
\begin{equation} \label{FRWmetric}
  d\tilde{s}^2 = \tilde{g}^{\mu \nu} dx_\mu dx_\nu = - dt^2 + a^2(t) d
  \Sigma^2,
\end{equation}
where $d\Sigma^2$ is the metric on surfaces of constant $t$, which are
assumed to be maximally symmetric ($S^3$, $\mathbb{R}^3$, or $H^3$.)
The vector field $A_a$, meanwhile, must simply be
\begin{equation} \label{FRWvector}
  A_a = A_t (dt)_a
\end{equation}
in order to respect the symmetry of the solution.  This later
condition is quite restrictive, since it implies that $F_{\mu \nu} =
2\partial_{[\mu} A_{\nu]} = 0.$  The equation of motion
\eref{full.NL.EOM.A} then simplifies drastically:
\begin{equation}
  \frac{1}{\mathcal{Q}^{3/2}} \left( \xi V(A^2) - 2 \mathcal{Q}
    V'(A^2) \right) = -2 \frac{d}{d(A^2)} \left[
    \frac{V(A^2)}{\sqrt{\mathcal{Q}}} \right] = 0. 
\end{equation}
In other words, the norm of the vector field $A$, as measured with
respect to the physical metric $g^{ab}$, is not found at the minimum
of the potential $V(A^2)$, but instead at the minimum of an effective
potential defined by
\begin{equation}
  V_\mathrm{eff} (A^2) = \frac{V(A^2)}{\sqrt{ 1 + \xi A^2}}.
\end{equation}
We can define $b^2$ such that $V'(-b^2) = 0$, and $\bar{b}^2$ such
that $V'_\mathrm{eff}(-\bar{b}^2) = 0$.  Note that in general, these two
quantities will differ.  For example, suppose the potential is of the
form $V(A^2) = \frac{\beta}{4} (A^2 + b^2)^2 + \Lambda$: a
``Higgs-like'' potential plus a cosmological constant term.  It is
then the case that
\begin{equation}
  - \bar{b}^2
  = \frac{1}{3 \xi } \left[ 2 \sqrt{ \left(1 - b^2 \xi
    \right)^2+\frac{3 \Lambda  \xi ^2}{\beta }} - b^2 \xi - 2 \right]
  = - b^2 + \frac{\xi \Lambda}{\beta} + \mathcal{O}(\xi^2).
\end{equation}

The equation of motion for the gravitational metric $\tilde{g}^{ab}$ is then
simply
\begin{equation}
  \tilde{G}_{ab} + 8 \pi G \tilde{\Lambda} \tilde{g}_{ab} = 0
\end{equation}
where $\tilde{\Lambda} \equiv V(-\bar{b}^2)$.  This implies that the
gravitational metric is de Sitter, anti-de Sitter, or Minkowski,
depending on the sign of $\tilde{\Lambda}$ and the Gaussian curvature
$\tilde{k}$ of the spatial hypersurfaces; the scale factor $a(t)$ will simply
obey the Friedmann equation
\begin{equation} \label{friedmann}
  \left( \frac{da}{dt} \right)^2 - \frac{8 \pi G \tilde{\Lambda}}{3} a^2
  = - \tilde{k}.
\end{equation}

To find the matter metric $g^{ab}$---which is, after all, what would
be measurable via observations of ``normal matter'' in such a
Universe---we first note that since
$A^2 = g^{ab} A_a A_b = - \bar{b}^2$, we have
\begin{equation}
  A_t = \frac{\bar{b}}{\sqrt{-g^{tt}}}.
\end{equation}
Recalling \eref{inverse.met.def}, this implies that
\begin{eqnarray}
  \tilde{g}^{\mu \nu} dx_\mu dx_\nu
  &= \frac{1}{\sqrt{1 - \xi
    \bar{b}^2} } \left(g^{\mu \nu} + \xi g^{\mu \rho} g^{\nu \sigma}
    A_\rho 
    A_\sigma \right) dx_\mu dx_\nu \nonumber \\
  &= \sqrt{1 - \xi \bar{b}^2} g^{tt} dt^2 + \frac{1}{\sqrt{1 - \xi
    \bar{b}^2}} g^{ij}(x^\mu) dx_i dx_j.
\end{eqnarray}
Comparing this to \eref{FRWmetric}, we conclude that
\begin{equation}
  ds^2 = g^{\mu \nu} dx_\mu dx_\nu = - \frac{dt^2}{\sqrt{1 - \xi
      \bar{b}^2}} + \sqrt{1 - \xi
    \bar{b}^2} a^2(t) d\Sigma^2.
\end{equation}
The constant in front of the spatial part of the metric can be
absorbed into $a(t)$, and we can rescale our time coordinate
$\bar{t} = t/(1 - \xi \bar{b}^2)^{1/4}$.  In terms of this coordinate,
the Friedmann equation \eref{friedmann} becomes 
\begin{equation} \label{friedmann.rescale} \left( \frac{da}{d\bar{t}}
  \right)^2 - \frac{8 \pi G \sqrt{1 - \xi \bar{b}^2}
    \tilde{\Lambda}}{3} a^2 = - \tilde{k} \sqrt{1 - \xi \bar{b}^2}.
\end{equation}
In other words, the matter metric is (like the gravitational metric)
de Sitter, anti-de Sitter, or Minkowski; however, the measurable
values of the cosmological constant $\bar{\Lambda}$ and the Gaussian
curvature $\bar{k}$ would be rescaled:
\begin{equation}
  \bar{\Lambda} = \tilde{\Lambda} \sqrt{1 - \xi \bar{b}^2}, \qquad
  \bar{k} = \tilde{k} \sqrt{1 - \xi 
    \bar{b}^2}.
\end{equation}

The effects of this model would therefore not be distinguishable from
a conventional FRW cosmology with $\Lambda \neq 0$; the net effect of
the non-trivial couplings between $A_a$ and the metric in
\eref{full.LV.NL.action.metric} is simply to rescale the ``bare''
values of the cosmological constant and the spatial curvature.

The lack of directly observable effects in this simplistic spacetime
does not necessarily imply that physically meaningful effects do not
exist in other circumstances.  The spatial isotropy of this spacetime
makes it a particularly poor test bed for the effects of a fundamental
non-zero vector field, since it implies that the field strength
vanishes identically. A more reasonable assumption in the context of
Lorentz symmetry violation would be a spacetime that is homogeneous
but anisotropic, with a local two-dimensional rotational symmetry at
every point corresponding to rotations keeping the spatial part of
$A_a$ fixed.  Metrics with this symmetry structure have been
previously examined in the context of perfect-fluid solutions
\cite{Kantowski1966,Ellis1969,King1973}, as well as in the context of
more recent vector-tensor models \cite{Heisenberg2016,
  Artymowski2012}.  In the Lorentz-violating bootstrap model, this
would lead to non-trivial dynamics for $A_a$, since we would now have
$F_{ab} \neq 0$; the equations of motion \eref{full.NL.EOM.A} and
\eref{full.NL.EOM.tg} would also become significantly more
complicated. Work on the evolution of such spacetimes in this model is
ongoing, and will be described in a future paper.

\ack

I would like to thank J.\ Tasson and S.\ Mukohyama for discussions
during the preparation of this work.  This research was supported in
part by Perimeter Institute for Theoretical Physics.  Research at
Perimeter Institute is supported by the Government of Canada through
Industry Canada and by the Province of Ontario through the Ministry of
Economic Development \& Innovation.

\section*{References}

\bibliographystyle{iopart-num-nourl}
\bibliography{Bootstrap_gravity}

\providecommand{\newblock}{}
\begin{thebibliography}{10}
\expandafter\ifx\csname url\endcsname\relax
  \def\url#1{{\tt #1}}\fi
\expandafter\ifx\csname urlprefix\endcsname\relax\def\urlprefix{URL }\fi
\providecommand{\eprint}[2][]{\url{#2}}

\bibitem{Will}
Will C~M 2014 {\em Living Rev. Relativ.\/} {\bf 17} 4 (\textit{Preprint}
  \eprint{1403.7377})

\bibitem{Colladay1998}
Colladay D and Kosteleck{\'{y}} V~A 1998 {\em Phys. Rev. D\/} {\bf 58} 116002

\bibitem{Kostelecky2008}
Kosteleck{\'{y}} V~A and Russell N 2011 {\em Rev. Mod. Phys.\/} {\bf 83} 11--31
  Current data tables available at arxiv:0801.0287.

\bibitem{Bailey2006}
Bailey Q and Kosteleck{\'{y}} V~A 2006 {\em Phys. Rev. D\/} {\bf 74} 045001

\bibitem{Kostelecky2015}
Kosteleck{\'{y}} V~A and Tasson J~D 2015 {\em Phys. Lett. B\/} {\bf 749}
  551--559 (\textit{Preprint} \eprint{1508.07007})

\bibitem{Kostelecky2016}
Kosteleck{\'{y}} V~A and Mewes M 2016 {\em Phys. Lett. B\/} {\bf 757} 510--514
  (\textit{Preprint} \eprint{1602.04782})

\bibitem{Bailey2016}
Bailey Q~G 2016 {\em Phys. Rev. D\/} {\bf 94} 065029

\bibitem{Kraichnan1955}
Kraichnan R~H 1955 {\em Phys. Rev.\/} {\bf 98} 1118--1122

\bibitem{Gupta1952}
Gupta S~N 1952 {\em Proc. Phys. Soc. London\/} {\bf A65} 608--619

\bibitem{Thirring1961}
Thirring W 1961 {\em Ann. Phys. (N. Y).\/} {\bf 16} 96--117

\bibitem{Bailey2004}
Bailey Q and Kosteleck{\'{y}} V~A 2004 {\em Phys. Rev. D\/} {\bf 70} 1--10

\bibitem{Kostelecky2011}
Kosteleck{\'{y}} V~A and Tasson J~D 2011 {\em Phys. Rev. D\/} {\bf 83} 1--59

\bibitem{Deser1970}
Deser S 1970 {\em Gen. Relativ. Gravit.\/} {\bf 1} 9--18

\bibitem{Kostelecky2009}
Kosteleck{\'{y}} V~A and Potting R 2009 {\em Phys. Rev. D\/} {\bf 79} 065018

\bibitem{Bluhm2017}
Bluhm R 2017 {\em Symmetry (Basel).\/} {\bf 9} 230

\bibitem{Kostelecky2004}
Kosteleck{\'{y}} V~A 2004 {\em Phys. Rev. D\/} {\bf 69} 105009

\bibitem{Iglesias2007}
Iglesias A, Kaloper N, Padilla A and Park M 2007 {\em Phys. Rev. D\/} {\bf 76}
  104001 (\textit{Preprint} \eprint{0708.1163})

\bibitem{Lanczos1957}
Lanczos C 1957 {\em Rev. Mod. Phys.\/} {\bf 29} 337--350

\bibitem{Flanagan2004}
Flanagan {\'{E}}~{\'{E}} 2004 {\em Class. Quantum Gravity\/} {\bf 21}
  3817--3829 (\textit{Preprint} \eprint{gr-qc/0403063})

\bibitem{Bluhm2003}
Bluhm R, Kosteleck{\'{y}} V~A, Lane C~D and Russell N 2003 {\em Phys. Rev. D\/}
  {\bf 68} 125008 (\textit{Preprint} \eprint{hep-ph/0306190})

\bibitem{Shao2018}
Shao C~G, Chen Y~F, Sun R, Cao L~S, Zhou M~K, Hu Z~K, Yu C and M{\"{u}}ller H
  2018 {\em Phys. Rev. D\/} {\bf 97} 24019

\bibitem{Bourgoin2017}
Bourgoin A, {Le Poncin-Lafitte} C, Hees A, Bouquillon S, Francou G and Angonin
  M~C 2017 {\em Phys. Rev. Lett.\/} {\bf 119} 1--6

\bibitem{Abbott2017}
Abbott B~P {\em et~al.\/} 2017 {\em Astrophys. J.\/} {\bf 848} L13

\bibitem{Heisenberg2014}
Heisenberg L 2014 {\em J. Cosmol. Astropart. Phys.\/} {\bf 2014} 015--015

\bibitem{Heisenberg2017}
Heisenberg L 2017 {\em Proceedings, 52nd Rencontres Moriond Gravit. (Moriond
  Gravit. 2017)\/} (\textit{Preprint} \eprint{1705.05387})

\bibitem{Heisenberg2016}
Heisenberg L, Kase R and Tsujikawa S 2016 {\em J. Cosmol. Astropart. Phys.\/}
  {\bf 2016}

\bibitem{Gumrukcuoglu2019}
G{\"{u}}mr{\"{u}}k{\c{c}}{\"{u}}oǧlu A~E and Koyama K 2019 {\em Phys. Rev.
  D\/} {\bf 99} 1--16

\bibitem{Kantowski1966}
Kantowski R and Sachs R~K 1966 {\em J. Math. Phys.\/} {\bf 7} 443

\bibitem{Ellis1969}
Ellis G~F~R and MacCallum M~A~H 1969 {\em Commun. Math. Phys.\/} {\bf 12}
  108--141

\bibitem{King1973}
King A~R and Ellis G~F~R 1973 {\em Commun. Math. Phys.\/} {\bf 31} 209--242

\bibitem{Artymowski2012}
Artymowski M and Lalak Z 2012 {\em Phys. Lett. B\/} {\bf 707} 203--208

\end{thebibliography}

\end{document}